\documentclass[aps,prl,showpacs,superscriptaddress,floatfix,twocolumn,notitlepage]{revtex4-1}

\usepackage{blindtext}
\usepackage{pdfsync}
\usepackage{graphicx}
\usepackage{float}
\usepackage[percent]{overpic}
\usepackage{amsmath,amssymb,amsfonts,mathrsfs}
\usepackage{bm}
\usepackage{bbold}
\usepackage{color}
\usepackage{ulem}
\usepackage{natbib}
\usepackage[us,12hr]{datetime} 
\usepackage{braket}
\usepackage{bbm}
\usepackage{subcaption}
\usepackage[breaklinks=true,colorlinks,citecolor=blue,linkcolor=blue,urlcolor=blue]{hyperref}
\begin{document}
\title{Floquet theory for the electronic stopping of projectiles in solids}

\author{Nicol\`o Forcellini}
\email{nf334@cam.ac.uk}
\affiliation{Theory of Condensed Matter, Cavendish Laboratory,
JJ Thomson Ave, Cambridge CB3 0HE, UK}
\author{Emilio Artacho}
\affiliation{Theory of Condensed Matter, Cavendish Laboratory,
JJ Thomson Ave, Cambridge CB3 0HE, UK}

\date{\today}

\begin{abstract}
  A general theoretical framework for the study of electronic stopping
of particle projectiles in crystalline solids is proposed.
  It neither relies on perturbative or linear response approximations, 
nor on an ideal metal host.
  Instead, it exploits the discrete translational symmetries in a space-time 
diagonal determined by a projectile with constant velocity moving along a 
trajectory with crystalline periodicity.
  This allows for the characterisation of (stroboscopically) stationary 
solutions, by means of Floquet theory for time-periodic systems.
  Previous perturbative and non-linear jellium models are recovered from 
this general theory.
  An analysis of the threshold velocity effect in insulators is 
presented based on Floquet quasi-energy conservation.
\end{abstract}

\maketitle


  Particles of radiation shooting through matter 
interact with the constituent nuclei and electrons and lose their kinetic 
energy to them. 
  The energy loss per unit length to the electrons (nuclei) is called 
electronic (nuclear) stopping power $S_{e(n)}$.
  It is of great applied interest in various contexts, since materials that 
can withstand ionic radiation have important applications for medical, nuclear 
and aerospace engineering industries.
  From the fundamental side it represents a paradigmatic problem in the context
of strongly non-equilibrium electronic systems.
  Electronic stopping of ions in solids has been studied for over a century.
  The most popular theoretical paradigm after the early works
\cite{bragg, bohr, bethe, bloch, fermiteller} is due to Lindhard 
in his linear-response theory of electronic stopping, of general applicability 
for any host material \cite{lindhard1,lindhard2}, and accessible to
first-principles theory \cite{Reining2016}.
  It assumes, however, a weak effective interaction between the projectile and 
the solid, which may be justified at very high velocities \cite{bethe}, but not 
in general.

  A fully non-linear theory for slow projectiles, with velocity $v$ much smaller than
the Fermi velocity of the electrons, was proposed for the homogeneous electron 
liquid (jellium) in the 70's by Ferrel and Ritchie \cite{ferrel} and then 
developed into a method for calculations by Enchenique, Nieminen and Ritchie 
\cite{echenique1}.
  It is based on the mapping of the electronic stopping power into the problem 
of electronic scattering by an impurity in the homogeneous electron liquid, 
when changing reference frame to the one where the projectile is stationary.
  A generalisation to arbitrary $v$ was developed later 
\cite{schonhammer,schonhammer2,zaremba}. 
  Although a very successful theory and paradigmatic reference for simple metals,
its extension to semiconductors, insulators, transition metals, etc. is 
very qualitative and limited. 

  Explicit simulations of the stopping process started over
a decade ago, in which a projectile moves within a solid in a large (periodic) 
simulation box, using either time-dependent tight-binding \cite{TB1,TB2,TB3} or 
first-principles time-dependent density-functional theory (TDDFT)
\cite{Pruneda2007, Krasheninnikov2007, Quijada2007, Hatcher2008, Correa2012,
Zeb2012, Ojanpera2014, Ullah2015, Wang2015, Schleife2015, Lim2016,
Quashie2016, Reeves2016, Li2017, Yost2017, Bi2017, Ullah2018}.
  They are computationally expensive, but allow the study of 
materials beyond simple metals, and have access \cite{Pruneda2007, Ullah2015, 
threshold3} to experimentally observed non-trivial effects such as the 
appearance of a threshold velocity \cite{markin,draxlerthreshold} for the 
onset of stopping in insulators and semiconductors.

  Those are by no means the only theories beyond linear response (see refs.
within \cite{sigmund1,sigmund2}).
  However, there appears to be no clear physical theory beyond the 
linear-response and jellium approximations that can account for the general 
properties of stopping for arbitrary crystalline systems, an issue that 
the novel framework presented in this letter intends to address.

\begin{figure}[H]
\centering
  \begin{subfigure}[b]{0.25\textwidth}
    \includegraphics[width=\textwidth]{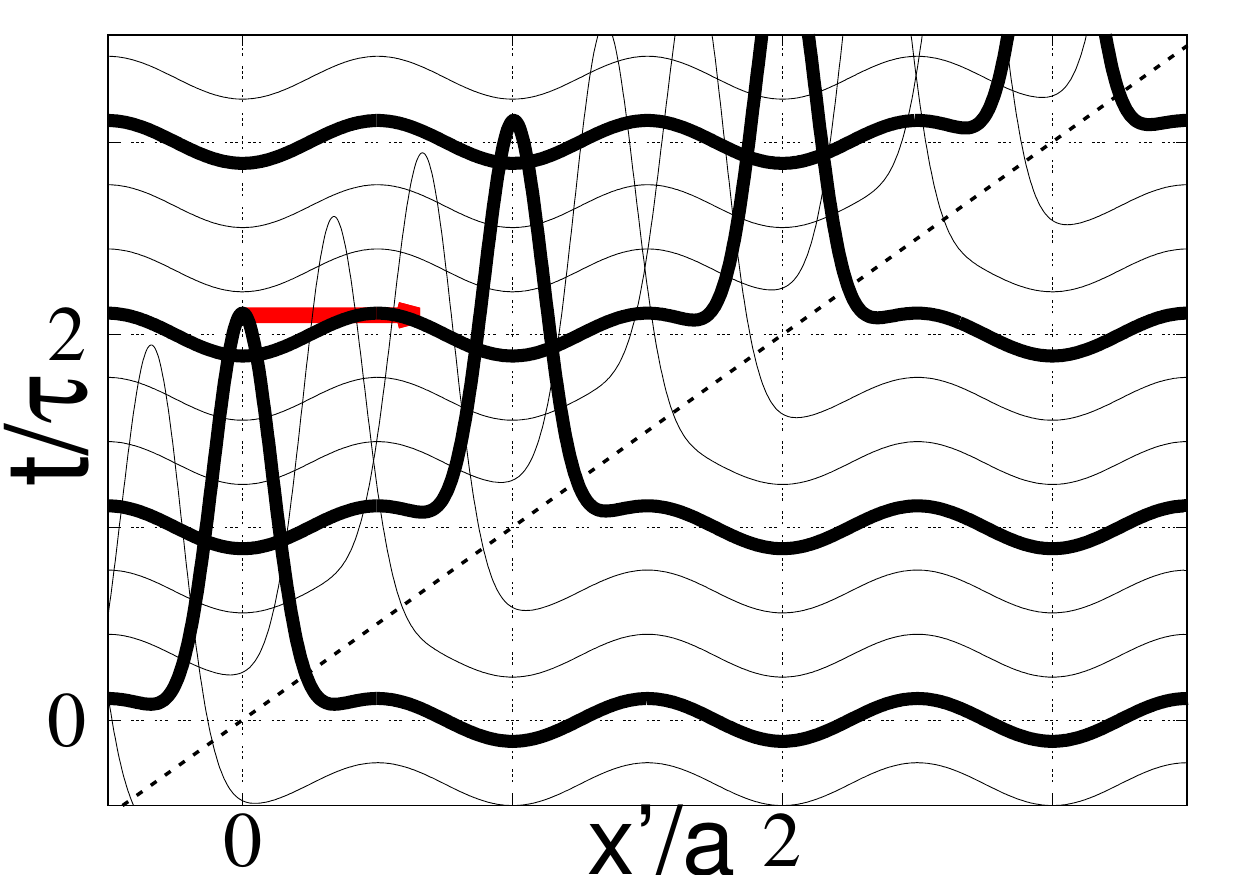}
    \label{fig:1}
  \end{subfigure}
  \hspace{-1.3em}%
  \begin{subfigure}[b]{0.25\textwidth}
    \includegraphics[width=\textwidth]{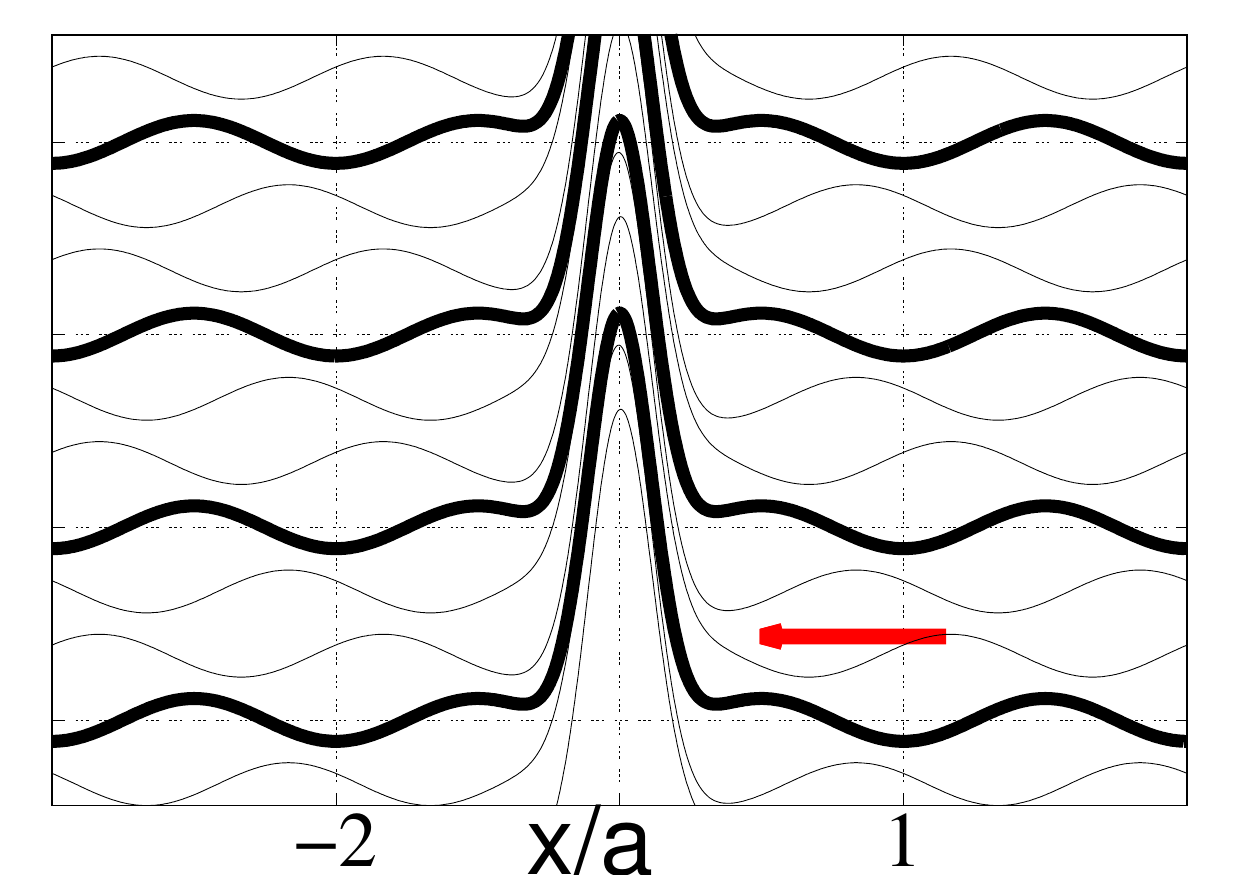}
    \label{fig:2}
  \end{subfigure}
\caption{Evolution of the sum of a crystalline and
a projectile potential in one dimension, both in the laboratory
reference frame (left) and in the projectile's (right).
  $a$ is the lattice parameter, $\tau=a/v$, where $v$ is the 
projectile's velocity (slope of dotted line).
  The curves depict potential versus $x$, and are shifted
for different times. Thicker lines indicate times separated by $\tau$.}
\label{figspace}
\end{figure}

\textit{Model}---
  The theory for jellium \cite{ferrel,echenique1,schonhammer,schonhammer2,zaremba} 
is implicitly built on the fact that the problem of a projectile of constant 
velocity $\mathbf{v} = v\hat{\mathbf{v}}$ moving in a homogeneous electron liquid, 
although a time-dependent, non-conservative problem, retains a continuous symmetry 
and related conservation, 
which neither stems from time nor space homogeneity, but rather from invariance along a 
space-time diagonal. 
  The change to the projectile's reference frame aligns this trajectory
with the time axis and the problem becomes energy conservative, while still 
dissipative in the laboratory frame.
  Consider the same projectile in a crystalline periodic solid, with a spatial 
periodicity $a$ along its trajectory.
  The translational invariance becomes discrete along the same line of space-time: 
the system is invariant under combined space-time translations 
${\cal{T}}^*: \mathbf{r} \rightarrow \mathbf{r}+na\hat{\mathbf{v}},\ t 
\rightarrow t+n\tau$ with $n$ integer, and $\tau=a/|\mathbf{v}|$.
  Changing to the projectile's reference frame the problem becomes 
purely time-periodic with period $\tau$, switching from ${\cal{T}}^*$-invariance 
to ${\cal{T}}: t \rightarrow t+\tau$ invariance.
  This implies that, not only the host electrons, but also the crystalline 
potential moves past the projectile with velocity $-v$,
which spoils the mapping used in refs.~\cite{ferrel, echenique1, schonhammer}.
  This is illustrated in Fig.~\ref{figspace}, and is the main point exploited 
in this work, as it allows a treatment based on Floquet theory for time-periodic 
Hamiltonians \cite{hanggi1, shirley}.

  A projectile with kinetic energy in e.g. the MeV scale will 
slow down while exciting the electron system at a rate of a few eV/\AA. 
  It is a strong, non-perturbative excitation, but the slowing down of 
the projectile is barely noticeable over a significant distance in the 
atomic scale.
  As done in the non-linear theory for jellium \cite{ferrel,echenique1,schonhammer,
schonhammer2,zaremba}, we consider the ideal non-conservative case of a 
projectile moving at a constant velocity along a rectilinear trajectory in a solid. 
  The following formalism will be limited to non-interacting particles, and the 
projectile will be represented by a local scalar potential.
  The method can be straightforwardly generalized to more realistic situations using 
time-dependent mean-field or Kohn-Sham methods to include realistic crystals, 
projectiles, and electron-electron interactions.

  A general time-dependent lattice-projectile model Hamiltonian, with the 
${\cal{T}}^*$ symmetry in the laboratory frame in which the lattice is at 
rest, can be written as 
\begin{equation}
H_{lab}(\mathbf{r'},t) = H_0(\mathbf{r'}) + V_P(\mathbf{r'},t),
\end{equation}
where $H_0(\mathbf{r'})$ is the lattice Hamiltonian and 
$V_P(\mathbf{r'},t) = V_P(\mathbf{r'-v}t)$ is the 
potential that describes a projectile with velocity $\mathbf{v}$.
  In the reference frame moving with the projectile, $\mathbf{r} = \mathbf{r'-v}t$ 
(primed/unprimed indices indicating lab/projectile reference frame, respectively)
the Hamiltonian becomes
\begin{equation}\label{eq:periodicham}
H(\mathbf{r},t) = H_0(\mathbf{r}+\mathbf{v}t) + V_P(\mathbf{r}),
\end{equation}
which is time-periodic with period $\tau = a/|\mathbf{v}|$.

\textit{Floquet theory.}---
  According to the Floquet theorem \cite{hanggi1, shirley}, for a time periodic Hamiltonian there are time-dependent 
solutions to the Schr\"odinger equation  of the form 
\begin{equation}\label{eq:floquetstates}
\psi_{\alpha}(\mathbf{r},t) = e^{-i\varepsilon_{\alpha}t/\hbar}\phi_{\alpha}(\mathbf{r},t),
\end{equation} 
where $\phi_{\alpha}(\mathbf{r},t)$, the \textit{Floquet mode}, has the same time 
periodicity of the Hamiltonian. 
  The real parameter $\varepsilon_{\alpha}$, the Floquet \textit{quasi-energy}, 
uniquely defines solutions up to multiples of $\hbar\omega, \ \omega = 2\pi/\tau$, 
and it is a conserved quantity. 
  Quasi-energies $\varepsilon_{p\alpha}= \varepsilon_{\alpha}+p\hbar\omega$ 
(integer $p$) are equivalent, since $\phi_{p\alpha}(\mathbf{r},t) = 
e^{ip\omega t}\phi_{\alpha}(\mathbf{r},t)$, both $\cal{T}$-periodic.
  The zeroth mode quasi-energy $\varepsilon_{0\alpha} = \varepsilon_{\alpha}$ can be 
defined on any interval of size $\hbar \omega$, e.g. $-\hbar \omega/2 < 
\varepsilon_{0\alpha} \leq \hbar \omega/2$, the equivalent of a 1$^{\text{st}}$ 
Brillouin Zone (BZ).
  The Floquet modes satisfy an eigenvalue 
equation in the quasi-energy
\begin{equation}\label{floquet}
{\cal H}(\mathbf{r},t)\phi_{p\alpha}(\mathbf{r},t) = \varepsilon_{p\alpha}
\phi_{p\alpha}(\mathbf{r},t),
\end{equation}
where the (Hermitian) Floquet Hamiltonian 
${\cal H}(\mathbf{r},t) = H(\mathbf{r},t) - i \hbar \frac{\partial}{\partial t}$ 
has been introduced, acting on states in the enlarged Hilbert space 
$\mathscr{H} \otimes \mathscr{T}$, where $\mathscr{T}$ is the space of 
time-periodic functions with period $\tau$, the inner product being
defined by \cite{hanggi1}
$\braket{\braket{f|g}} = \frac{1}{\tau} \int_0^{\tau}
 dt \int_{} d\mathbf{r} \ f^*(\mathbf{r},t) g(\mathbf{r},t)$,
for complex functions $f(\mathbf{r},t)$ and $g(\mathbf{r},t)$.

  A first important consequence of this theory
is that the Floquet modes in the projectile's frame define the 
stationary solutions to the stopping problem in the lab frame.
  In previous theoretical work \cite{lindhard2,echenique1,schonhammer}, 
stationary solutions were either assumed or a direct consequence of key 
approximations.
  Their existence and character appear now naturally from Floquet theorem.
  Stationary now means $\cal{T}$-periodic, or \textit{stroboscopic}, i.e.
time-independent if looking at it at instants $t=t_0 + n\tau$ for 
$n\in\cal{Z}$.
  It does not mean these are the only expected solutions.
  In addition to transients related to occasional perturbations, 
one can also foresee deviations like the flapping instability
recently proposed \cite{Ullah2018}, which represents the analog
of a charge density wave along the $\cal{T}^*$-symmetric direction 
in space-time, a generalization of the time-crystal idea.


\textit{Stopping from Bloch-Floquet scattering theory}.---
  The stopping problem in the lab frame becomes a scattering one for the 
Floquet modes in the projectile's, in analogy with the theory for jellium,
replacing energy conservation by quasi-energy conservation and 
treating time $t$ as just an additional degree of freedom at the same 
level of a spatial coordinate \cite{martinez,sambe}.
  The asymptotic scattering states away from the projectile consist of the 
Bloch states of the crystal transformed to the projectile's frame.
  The Floquet Hamiltonian for the lattice-projectile system is defined as 
${\cal H}(\mathbf{r},t) = {\cal H}_0(\mathbf{r},t)+V(\mathbf{r})$, 
where ${\cal H}_0(\mathbf{r},t) = H_0(\mathbf{r+v}t)-i\hbar\frac{\partial}{\partial t}$ 
is the \textit{lattice} Floquet Hamiltonian, periodic with period $\tau$, whose 
complete set $\{\phi_{\alpha}(\mathbf{r},t)\}$ of eigenmodes are readily extracted 
from the Bloch states in the lab frame 
$\psi'_{n\mathbf{k}}(\mathbf{r'}) = e^{i\mathbf{k}\cdot\mathbf{r'}}
u_{n\mathbf{k}}(\mathbf{r'})$, with energy $E_{n}(\mathbf{k})$ and band index $n$,
which, transformed to the projectile's frame, become \cite{landau}
\begin{equation}\label{eq:blochfloquet}
\psi_{n,\mathbf{k}}(\mathbf{r},t) = u_{n\mathbf{k}}(\mathbf{r}+
\mathbf{v}t)e^{i(\mathbf{k}-m\mathbf{v}/\hbar)\cdot\mathbf{r}}
e^{-i\varepsilon_{n}(\mathbf{k}) t/\hbar},
\end{equation}
where $\varepsilon_{n}(\mathbf{k}) = E_n(\mathbf{k}) -\hbar\mathbf{k}\cdot\mathbf{v} + 
mv^2/2$, and $m$ is the electron mass.
  By comparing with Eq.~\ref{eq:floquetstates}, the quasi-enegies and Floquet modes 
are immediately identified as $\varepsilon_{\alpha} = \varepsilon_{n}(\mathbf{k})$ 
and $\phi_{\alpha}(\mathbf{\xi})  = \phi_{n\mathbf{k}}(\mathbf{r},t) 
\equiv u_{n\mathbf{k}}(\mathbf{r}+\mathbf{v}t)e^{i(\mathbf{k}-m\mathbf{v}/\hbar)\cdot\mathbf{r}}$
(Bloch-Floquet modes henceforth).
  The Floquet BZ for quasi-energy can be chosen to 
coincide with the BZ for the Bloch vectors: shifting $\mathbf{k}$ by 
$p \mathbf{G}_0$ (for $p\in\cal{Z}$ and $\mathbf{G}_0 = (2\pi/a)\mathbf{\hat{v}}$)
shifts the  quasi-energy by $p\hbar\omega$.

  Consider an initial Bloch state in the moving frame 
$\psi_{n\mathbf{k}_i}(\mathbf{r},t) = e^{-i\varepsilon_n(\mathbf{k}_i)t/\hbar}
\phi_{n\mathbf{k}_i}(\mathbf{r},t)$: 
adding the projectile, a perturbation of arbitrary strength which does not break 
the symmetry ${\cal{T}}$, the periodic mode of the full solution 
$\Psi^{(\pm)}_{n\mathbf{k}_i} = e^{-i\varepsilon_n(\mathbf{k}_i)t/\hbar}
\Phi^{(\pm)}_{n\mathbf{k}_i}(\mathbf{r},t)$ 
with quasi-energy $\varepsilon_{n}(\mathbf{k}_i)$ can be expressed as an integral 
equation in the Lippmann-Schwinger spirit, with $\mathbf{\xi} = (\mathbf{r},t)$,
\begin{equation}\label{eq:exactsol}
\Phi^{(\pm)}_{n\mathbf{k}_i}(\mathbf{\xi}) = \phi_{n\mathbf{k}_i}(\mathbf{\xi}) 
+ \int d\mathbf{\xi''} {\cal{G}}^{(\pm)}_{0}(\varepsilon_{n}(\mathbf{k}_i)|{\xi,\xi''})
V_P(\mathbf{r}'')\Phi^{(\pm)}_{n\mathbf{k}_i}(\mathbf{\xi}''),
\end{equation}
where the $(\pm)$ sign indicates outgoing/incoming boundary conditions, 
and an averaging over one cycle time $t''$ is implied. 
  ${\cal{G}}^{\pm}_{0}(\varepsilon_n(\mathbf{k}_i)|{\xi,\xi''})$ is the propagator 
for ${\cal H}_0$, which using completeness of the Floquet modes is \cite{martinez,economou}
\begin{equation}\label{eq:propagator}
{\cal{G}}^{(\pm)}_{0}(\varepsilon |{\xi,\xi''}) = \sum_p  \int d\beta 
\frac{\phi_{p\beta}(\mathbf{\xi})\phi^*_{p\beta}(\mathbf{\xi}'')}{\varepsilon
-\varepsilon_{p\beta} \pm i\eta}.
\end{equation}

\begin{figure}[H]
\centering
\includegraphics[width=8.cm]{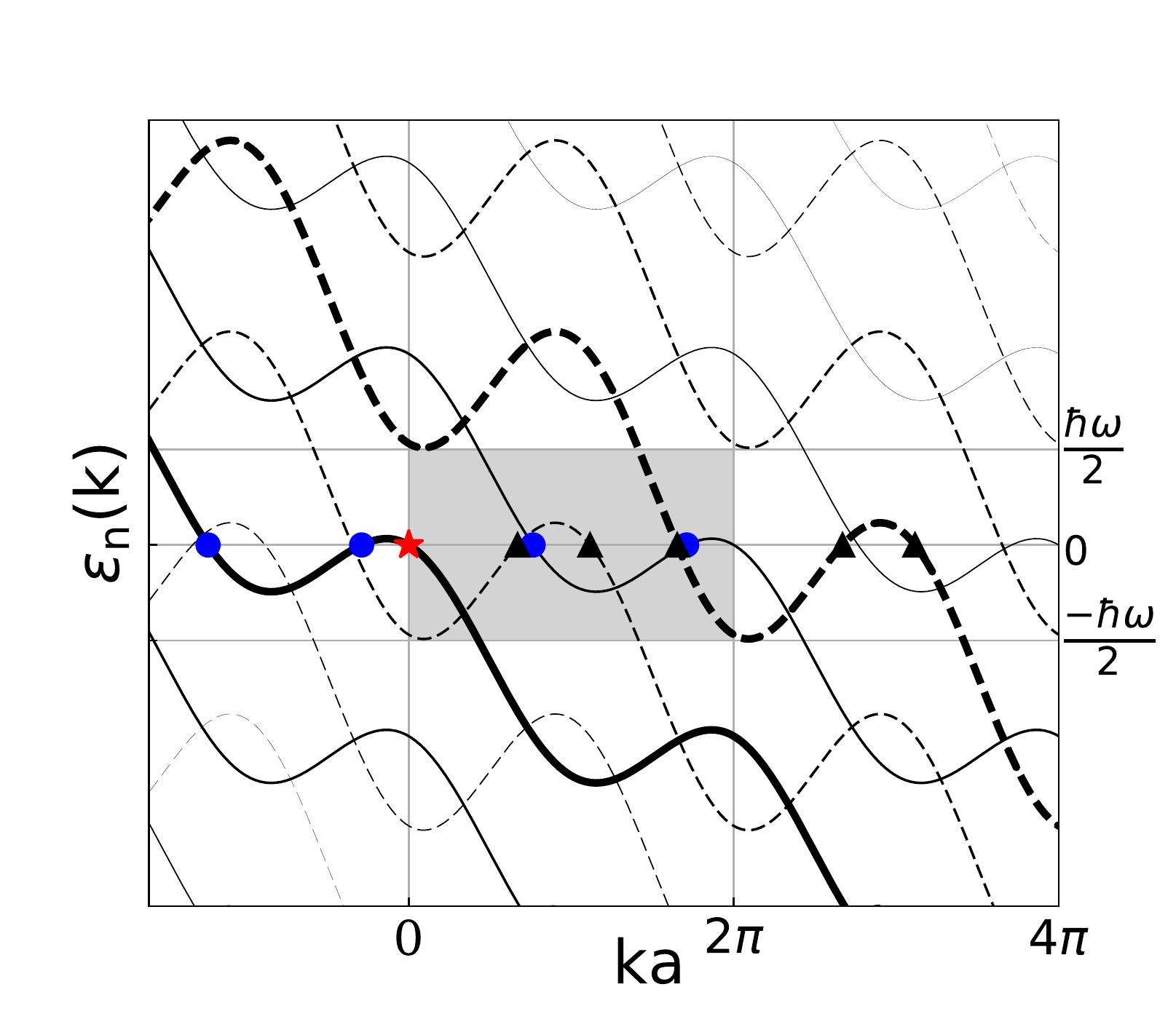}
\caption[justification=justify]{Example of quasi-energies $\varepsilon_{n}(k) = E_n({k}) -
\hbar{k}{v}+1/2mv^2$ of the Bloch-Floquet modes of a one-dimensional two-band 
system (thick lines), extended zone scheme and Floquet replicas of quasi-energy 
$\varepsilon_{pn}(k) = \varepsilon_{n}(k) +p\hbar\omega$ included (thin lines), 
$v > 0$.
  A choice for the 1$^{\text{st}}$ BZ for Bloch-Floquet modes is highlighted 
in grey. 
  Point and triangle markers correspond to solutions of the quasi-energy conservation 
equation for each band for an incoming mode at $k=0$ (red star).}
\label{figscattering1D}
\end{figure}

  Let us start from a general 1D system for simplicity, where $\xi = (x,t)$.
  The asymptotic behaviour of Eq.~\ref{eq:exactsol} in terms of the incoming
and outgoing Bloch-Floquet modes and the scattering amplitudes is 
\begin{equation}\label{eq:asymptotic}
\Phi^{(+)}_{nk_i}(x,t) \sim \phi_{nk_i}(x,t) + \sum_{m,k_f}A_{nk_i,mk_f} \phi_{mk_f}(x,t).
\end{equation}
  The band index $m$ and momenta $k_f$ of the scattered states are determined by the 
quasi-energy conservation condition $\varepsilon_n(k_i) = \varepsilon_m(k_f)$, which 
has in general multiple solutions --see Fig.~\ref{figscattering1D} for an example.
  It can also be expressed, the explicit expression for the quasi-energy from Eq.~\ref{eq:blochfloquet} 
(and using 3D notation) as
\begin{equation}\label{eq:qecons}
E_{m}(\mathbf{k}_f)-E_n(\mathbf{k}_i)  = \hbar(\mathbf{k}_f-\mathbf{k}_i)
\cdot\mathbf{v}\; .
\end{equation}
  This expression, appearing naturally from 
quasi-energy conservation, coincides with what obtained from energy and 
momentum conservation in a collision of an electron with a projectile of 
mass $M_P \rightarrow \infty$ \cite{Ullah2015}, and in perturbation theory 
\cite{emilio2007}. 
  The possible values of the $\mathbf{k}$'s in Eq.~\ref{eq:qecons} are not limited 
to the 1st BZ, but must be considered in the extended zone 
scheme as in Fig.~\ref{figscattering1D}, or, equivalently, on bands shifted 
by a multiple of $\hbar\omega$.
The scattering state needs to fulfill outgoing boundary conditions, i.e. 
the group velocity defined as $v_g^m(k_f) = \frac{1}{\hbar}\frac{\partial}{\partial k} 
(\varepsilon_{m}(k))|_{k = k_f} = \frac{1}{\hbar} 
\frac{\partial}{\partial k}E_n(k) - v$ has to point away from the projectile.
  The state in Eq.~\ref{eq:asymptotic} differs from the 1D free-particle case, whose asymptotic scattering
states are plane waves \cite{echenique1,schonhammer,schonhammer2} with only two 
outgoing channels per particle (reflected and transmitted) from energy 
conservation, similarly to Bloch-wave scattering by defects and 
impurities \cite{newton,korringa,kohn}.
  From Eq.~\ref{eq:exactsol}, the scattering amplitudes in Eq.~\ref{eq:asymptotic} are
\begin{equation}\label{eq:scattampl}
A_{nk_i,mk_f}= -\frac{i}{\hbar|v_g^m(k_f)|}
\braket{\braket{ \phi_{mk_f}(t)|V_P|\Phi^{(+)}_{nk_i}(t)}}.
\end{equation}

  Outgoing scattering states have in general different group velocity and satisfy
\begin{equation}
\sum_{mk_f} |A_{nk_i,mk_f}|^2|v_g^m(k_f)| = |v_g^n(k_i)|,
\end{equation}
from conservation of probability flux. 
  From this relation an expression for the energy transfer rate (ETR) 
to the electron system in the lab frame can be derived. 
  For the single-particle scattering state of Eq.~\ref{eq:asymptotic}, considering the 
energy flux difference between the outgoing and incoming states one gets
\begin{equation}\label{eq:etr}
\begin{split}
&\dot{E}_{ni} = \rho_i\sum_{mk_f}\left( |A_{nk_i,mk_f}|^2|v_g^m(k_f)|E_{m}(k_f) - 
|v_g^n(k_i)| E_{n}(k_i) \right) \\
& =  -\rho_i \frac{2\pi}{\hbar}\sum_m \int \frac{dk_f}{2\pi} (E_m(k_f)-E_n(k_i))
\times \\
&|\braket{\braket{ \phi_{mk_f}|V_P|\Phi^{(+)}_{nk_i}}}|^2
\delta(E_p(k_f)-E_n(k_i)-v\hbar(k_f-k_i)),
\end{split}
\end{equation}
where $\rho_i$ corresponds the density of the incoming state.
The corresponding expression in 3D is 
\begin{equation}\label{eq:etr3D}
\begin{split}
&\dot{E}_n(\mathbf{k}_i) = \rho_i\frac{2\pi}{\hbar} \sum_m \int  
\frac{d^3k_f}{(2\pi)^3}\Delta E_{mn,fi} \\
&\times |\braket{\braket{ \phi_{m\mathbf{k}_f}|V_P|\Phi^{(+)}_{n\mathbf{k}_i}}}|^2 
\delta(\Delta E_{mn,fi}-\mathbf{v} \cdot\hbar\Delta\mathbf{k}_{fi}),
\end{split}
\end{equation}
where $\Delta E_{mn,fi}= E_{m}(\mathbf{k}_f)- E_{n}(\mathbf{k}_i)$, 
$\Delta\mathbf{k}_{fi} = \mathbf{k}_f-\mathbf{k}_i$.
  It is important to note here that the ETR is averaged over 
a period $\tau$, as the scattering amplitude coefficients are proportional to 
the the (time-averaged) matrix element $\braket{\braket{V_P}}$.
  Electronic stopping can be defined as $S_e =\dot{E}/v$, where $\dot{E}$ is 
here the total ETR, to be calculated by considering
the contributions of all of the possible transitions between occupied and unoccupied 
states. At temperature $T=0$, assuming occupied bands $n$ and unoccupied bands 
$m$ and integrating separately over initial and final momenta, it is
\begin{equation}\label{eq:stopping}
\begin{split}
&S_e = \frac{1}{v} \sum_{nm}\frac{2\pi}{\hbar}\int \frac{d\mathbf{k}_i}{(2\pi)^3} 
\int \frac{d{\mathbf{k}_f}}{(2\pi)^3}\Delta E_{mn,fi} \\
&\times |\braket{\braket{ \phi_{m\mathbf{k_f}}|V_P|\Phi^{(+)}_{n\mathbf{k_i}}}}|^2 
\delta(\Delta E_{mn,fi}-\mathbf{v} \cdot\hbar\Delta\mathbf{k}_{fi}),
\end{split}
\end{equation}
where the 3D version of Eq.~\ref{eq:scattampl} was used.
  To extend this to $T \neq 0$ and partially filled bands the relevant occupation 
numbers need to be introduced.

\begin{figure}[H] 
\centering
\begin{subfigure}[b]{0.85\linewidth}
  \centering
   \includegraphics[width=\textwidth]{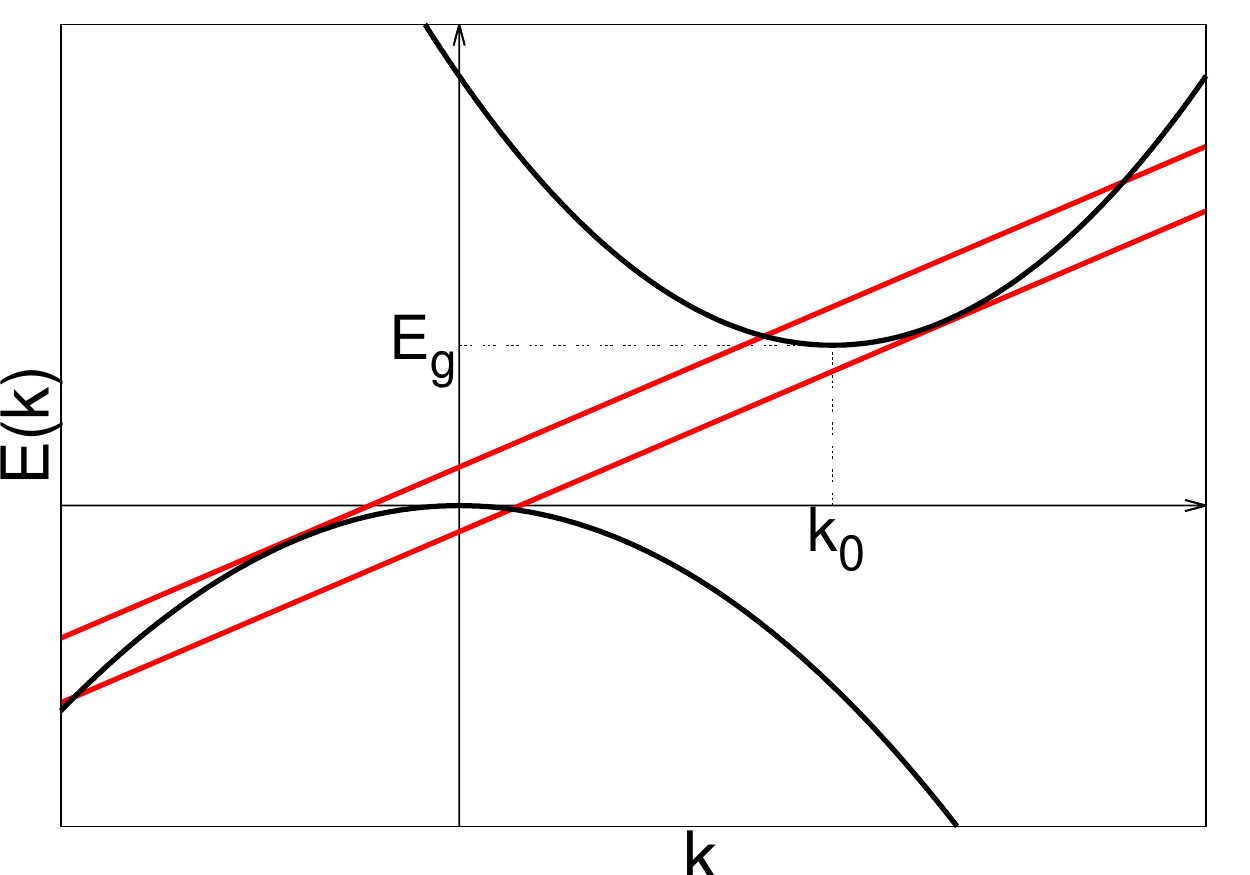}  
\end{subfigure}
\vspace{-0.2cm}
\begin{subfigure}[b]{0.98\linewidth}
  \centering
   \includegraphics[width=\textwidth]{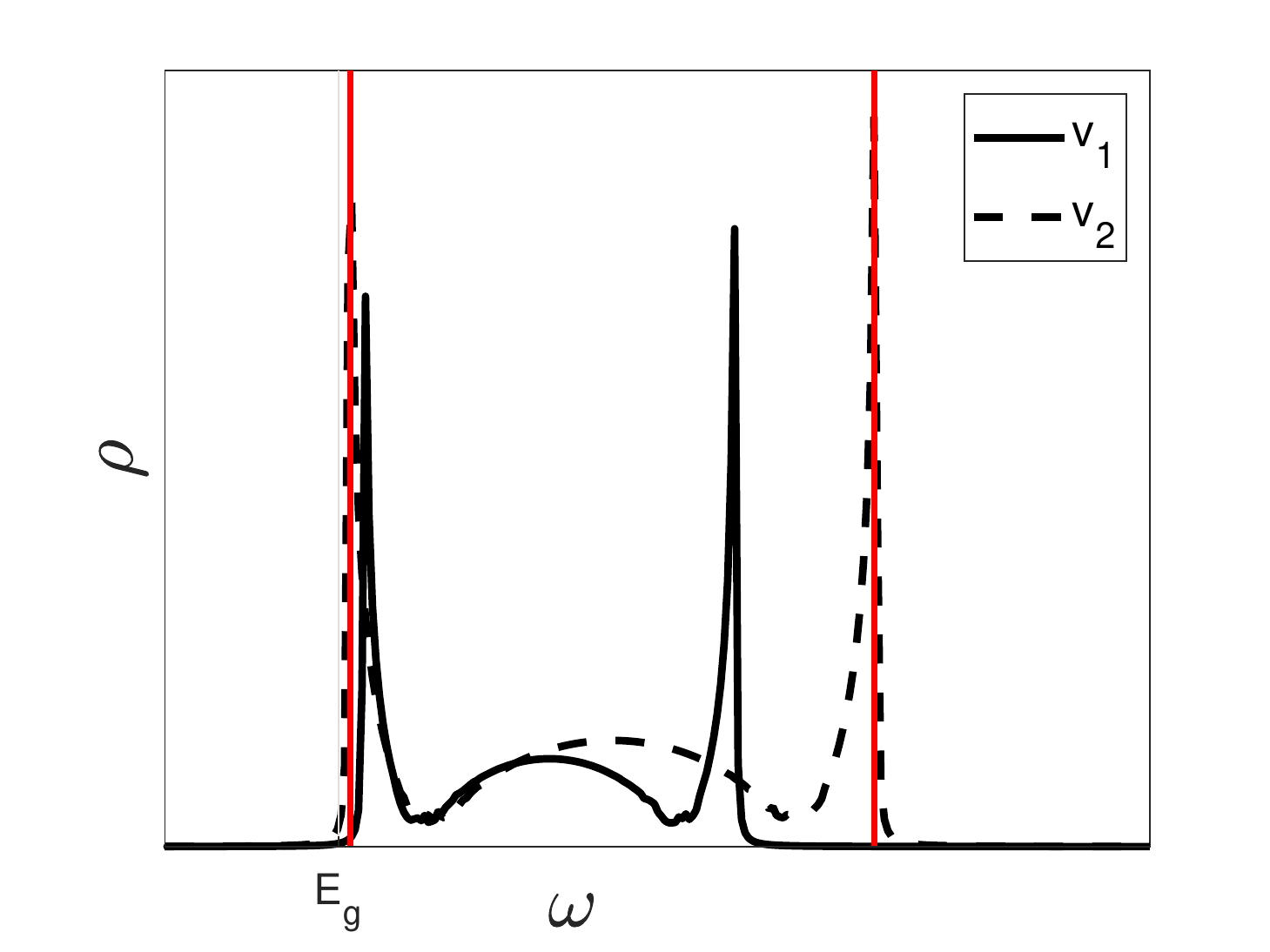}  
\end{subfigure}
\caption{Model one-dimensional insulator with parabolic bands and 
an indirect band gap (above). Red lines delimit range of possible 
electron-hole pair transitions compatible with Eq.~\ref{eq:qecons} 
for projectile velocity $v$ defining their slope.
  Below: Joint density of states (JDOS) $\rho(\omega,v)$ versus excitation
energy $\omega$ plotted for velocities $v_2 > v_1  > v_{th}$ for the same 
model. 
  The spectral limits for $v_2$ show the van Hove singularities related 
to the red lines above (outer tails due to artificial broadening).}
\label{fig:indirect}
\end{figure}

  These are very general expressions, which can be related to previous results 
and theoretical models.
  The homogeneous electron liquid theory 
\cite{ferrel,echenique1,schonhammer,schonhammer2,zaremba} is recovered 
from Eq.~\ref{eq:etr3D} for the ETR and from Eq.~\ref{eq:stopping} for the stopping 
power.
  Alternatively, if the projectile is treated as a small perturbation, the 
equivalent of a 1$^{\text{st}}$ Born approximation for Floquet scattering can
be used \cite{bilitewski}: substituting $\ket{\Phi^{(+)}_{nk_i}}$ by 
$\ket{\phi_{nk_i}}$ and assuming a smooth projectile $V_P(\mathbf{r})$, 
the scattering amplitude matrix elements in Eq.~\ref{eq:scattampl} become
\begin{equation}\label{eq:pertsol}
\braket{\braket{ \phi_{mk_f}|V_P|\phi^{}_{nk_i}}} \propto  \tilde{V}_P
(\Delta \mathbf{k}),
\end{equation}
where $\Delta \mathbf{k} = |\mathbf{k}_f- \mathbf{k}_i|$, and
$\tilde{V}_P(\Delta \mathbf{k})$ indicates the Fourier transform of 
$V_P(\mathbf{r})$, thereby recovering perturbation theory results
(see e.g. \onlinecite{emilio2007}).


\textit{Threshold velocity for insulators}.--- 
  The low-$v$ limit for the stopping of ions in a gapped material is now analyzed, 
which is not well described by earlier theories, and proved to be quite 
controversial in experiments \cite{moller, markin, draxlerthreshold}.
  We do it for a model insulators with parabolic energy bands around
the gap and isotropic effective masses $m_e$ and $m_h$ for electrons and
holes. 
  Let us consider an indirect band gap $E_g$ with the bottom of the 
conduction band displaced by $\mathbf{k}_0$ from the top of the valence
band, and a projectile travelling with velocity $\mathbf {v}$ parallel to 
$\mathbf{k}_0$.
  A joint density of states (JDOS) can be defined in analogy with optical 
transitions \cite{madelung}, which offers interesting insights 
(see Fig.~\ref{fig:indirect}). 
  Stopping power $S_e(v)$ follows by suitable integration in energy of the JDOS.

\begin{figure}[H]
\centering
\begin{subfigure}[b]{0.88\linewidth}
  \includegraphics[width=\textwidth]{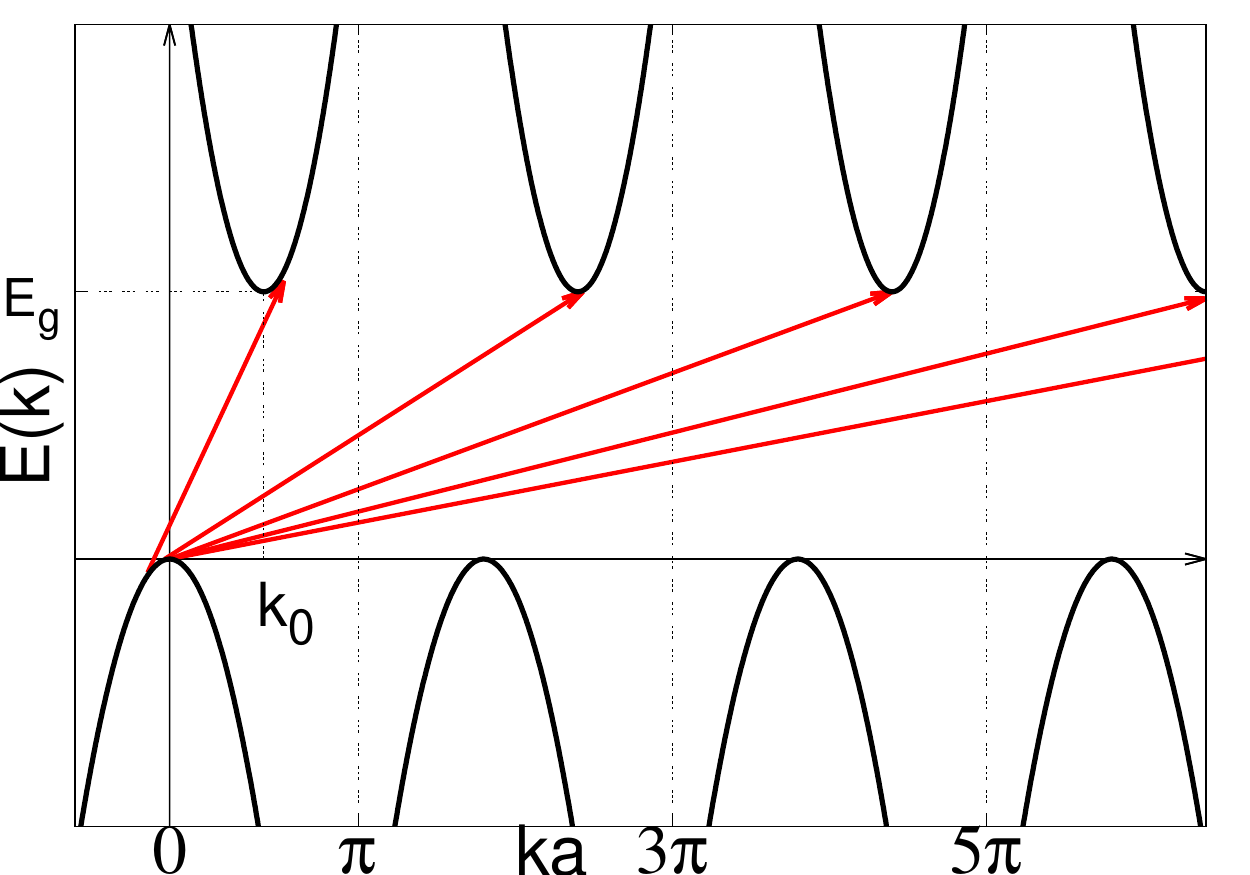}  
  \label{fig:sub-third}
\end{subfigure}
\hspace{0.5em}
\begin{subfigure}[b]{0.87\linewidth}
  \centering
  \includegraphics[width=\textwidth]{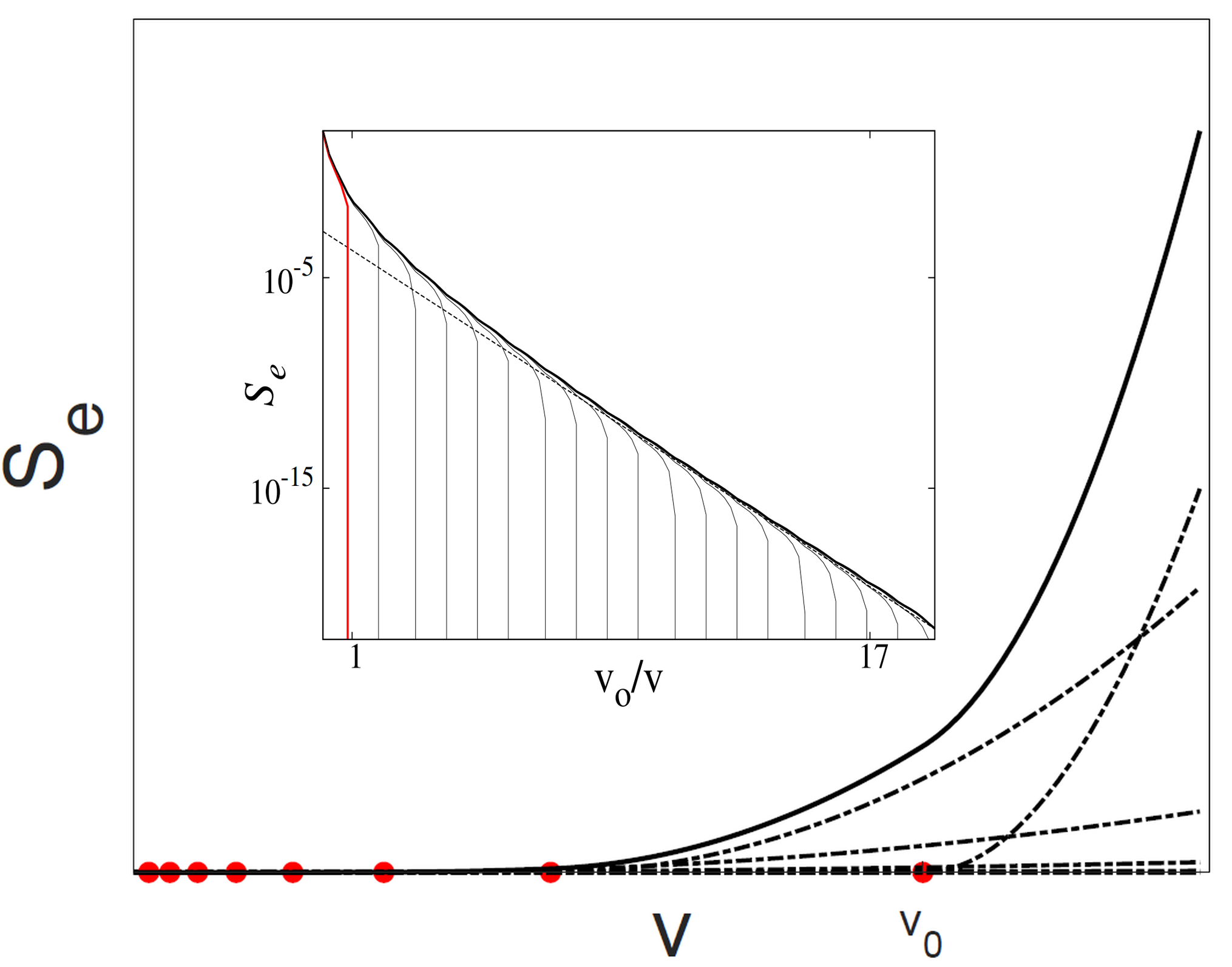}  
  \label{fig:sub-fourth}
\end{subfigure}
\label{fig:results}
\caption{Top: Partial threshold velocities (slopes of red lines, $v^p_{th}$) 
for the different replicas of parabolic bands in the extended zone scheme, 
corresponding to shifted Floquet modes, as in Fig.~\ref{figscattering1D}.
  Bottom: effective threshold behaviour for $S_e$ versus $v$ 
in  the small $v$ limit, for a 3D indirect gap model with 
$\gamma_l \propto e^{-\alpha l^{\lambda}}$.
Red dots correspond to first values of $v_{th}^{(p)}$, slopes of red arrows 
in top panel. 
 $v_0$ is the threshold velocity for transitions within the 1st BZ. 
  Inset: $\log S_e$ vs $1/v$.} 
\label{fig:repeatedindirect}
\end{figure}

  For the parabolic model in Fig.~\ref{fig:indirect} no stopping is
allowed below a threshold velocity $v_{th}$.
  For an actual insulator, however, the threshold behavior is less clean. 
  In fact, the adiabatic limit $v \rightarrow 0$ is quite non-trivial, as illustrated
in Fig.~\ref{fig:repeatedindirect}:
by quasi-energy conservation (Eq.~\ref{eq:qecons}) transitions are allowed for 
arbitrarily small $v$ even for gapped solids.
  This is shown in the figure using the repeated zone scheme implied in 
Eq.~\ref{eq:qecons}, where the lines of allowed transitions 
decrease in slope with decreasing $v$.
  Importantly, this picture is general and independent of 
perturbation theory \cite{emilio2007,elevator2}.
  The $S_e(v)$ curve is characterized by a series of onset
velocities, or partial thresholds, $v^{(p)}_{th}$, for $p\in {\cal Z}_{\ge 0}$  
(slopes of red lines and red dots in the upper and lower panels of 
Fig.~\ref{fig:repeatedindirect}, respectively), defined by
\begin{equation}
E_g = \frac{1}{2}(m_e+m_h)(v^{(p)}_{th})^2+\hbar(k_0 +\frac{2\pi}{a}p) 
v^{(p)}_{th} \; .
\end{equation}
  In the low-$v$ limit (large $p$), 
$v^{(p)}_{th}\sim \frac{E_g}{2\pi/a} \frac{1}{p}$.
  The stopping power for $v_{th}^{(p)}$ can be now approximated as
\begin{equation}\label{eq:stoppinglowv}
    S_e(v^{(p)}_{th}) \approx \sum_{l=p}^{\infty} 
    \gamma_l \, \, f(v_{th}^{(p)}-v_{th}^{(l)})
\end{equation}
summing over all replicas $l$ beyond the $p$-th, where
$\gamma_l$, relating to the scattering rate for transitions to
the $l$-th replica, is taken to decay with $l$, and assumed
constant for states within the $l$-th parabola,
and where $f(v-v_{th}^{(l)})$ is the stopping power contribution of the 
$l$-th replica.
  Close to each replica onset, $f(v) \propto v^m $, where $m$ depends on
dimension ($m=1$ in 1D, $m=2$ in 3D).
  Assuming an algebraic decay, $\gamma_l\sim l^{-\mu}$, 
\begin{equation}
S_e(v^{(p)}_{th}) \approx S_0 \sum_{l=p}^{\infty}
\left ( \frac{1}{p}-\frac{1}{l} \right )^m \frac{1}{l^{\mu}}
\end{equation}
giving the low-$v$ behaviour $S_e(v) \sim v^{m+\mu-1}$. 
  For a quicker decay, $\gamma_l \sim e^{-\alpha l^{\lambda}}$, then
$S_e(v) \sim e^{-(v^*/v)^{\lambda}}$.

  In summary, the presented Floquet theory provides a natural 
framework for the description of the stroboscopic stationary states 
arising in electronic stopping processes, as well as the reference
states for possible instabilities along the space-time symmetric
direction, analogous to CDWs in space, or time crystals in time.
  A general expression for the electronic stopping power has been derived 
and analysed.
  Previous perturbative \cite{emilio2007,bilitewski} and non-linear jellium 
\cite{echenique1,schonhammer,schonhammer2} theories are recovered, in the
limits of either weak coupling or homogeneous electron host.
  Floquet quasi-energy conservation has allowed the characterization of 
velocity thresholds in insulators, which prove to be far from trivial.
  The theory provides a paradigm for the understanding of electronic stopping
processes, and should lead to predictive computational schemes possibly 
more efficient than today's.



\end{document}